\documentclass[prl,twocolumn,aps,a4paper,superscriptaddress]{revtex4}
\usepackage[utf8]{inputenc} % set input encoding (not needed with XeLaTeX)

\usepackage{graphicx} % support the \includegraphics command and options

\usepackage{braket} 
\usepackage{amsmath} 
\usepackage{amssymb} 
\begin{document}

\title{A single photon transistor based on superconducting systems}
\author{Marco T. Manzoni}
\affiliation{QUANTOP -- Danish Quantum Optics Center, The Niels Bohr Institute, University of Copenhagen, Blegdamsvej 17, DK-2100 Copenhagen \O, Denmark}
\affiliation{Dipartimento di Fisica, Universit\`a degli Studi di Milano, I-20133 Milano, Italy}
\affiliation{ICFO-Institut de Ciencies Fotoniques, 08860 Castelldefels (Barcelona), Spain} 
\author{Florentin Reiter}
\affiliation{QUANTOP -- Danish Quantum Optics Center, The Niels Bohr Institute, University of Copenhagen, Blegdamsvej 17, DK-2100 Copenhagen \O, Denmark}
\author{Jacob Taylor}
\affiliation{Joint Quantum Institute/National Institute of Standards and Technology, 100 Bureau Dr MS 8410, Gaithersburg, MD  20899}
\author{Anders S. S\o rensen}
\affiliation{QUANTOP -- Danish Quantum Optics Center, The Niels Bohr Institute, University of Copenhagen, Blegdamsvej 17, DK-2100 Copenhagen \O, Denmark}

%\date{} % Activate to display a given date or no date (if empty),
% otherwise the current date is printed 

\begin{abstract}
We present a realistic scheme for how to construct a single-photon transistor where the presence or absence of a single microwave  photon controls the propagation of a subsequent strong signal signal field. The proposal is designed to work with existing superconducting artificial atoms coupled to cavities. We study analytically and numerically the efficiency and the gain of our proposal and show that current transmon qubits allow for error probabilities $\sim$ 1$\%$ and gains of the order of hundreds. 
\end{abstract}

\maketitle

In analogy with electronic transistors, a single photon transistor is a device where the presence or absence of a single gate photon controls the propagation of a large number of signal photons \cite{transistor,transistorexp}. Such devices would represent a milestone  enabling a plethora of new approaches for processing light, but their realization is hampered by the absence of interactions between photons.  A promising route towards strong interactions at the single photon level consists of coupling propagating photons to individual atom-like systems \cite{akimov, claudon, lee, wilk, Volz,Englund,maiwald}. The best realization of such a coupling is  achieved in the microwave regime where experiments have demonstrated an unprecedented control of the coupling between superconducting artificial atoms and microwave photons \cite{superconducting_review, circuit_qed}. In this letter we describe how to realize a single photon transistor based on existing superconducting technology. The resulting devices can be directly employed to detect individual itinerant microwave photons, and may find a range of applications within quantum information processing.

Various protocols for single photon non-linearities, quantum gates, and transistors using superconducting systems have been proposed \cite{rebic,neumeier,adhikari,dicarlo}, but have relied on unconventional qubit designs, multiple qubits and isolators, or temporal switching of
parameters.
A different method to construct the desired single photon transistors %based on the coupling of traveling photons in a one dimensional waveguide
was described in Ref. \cite{transistor}, and realized in the optical regime in Ref. \cite{transistorexp}. The atomic level structure used in that work is, however, not compatible with current superconducting artificial atoms (for convenience these will be referred to as atoms below). Here we show how to realize this proposal for a three level ladder ($\Xi$) system coupled to a superconducting cavity. This is the generic level scheme for the so-called transmon \cite{koch,fedorov,rigetti,sears,houck} and phase qubits \cite{neeley} which essentially constitutes anharmonic ladders. 

\begin{figure}[h]
%\centering
\includegraphics{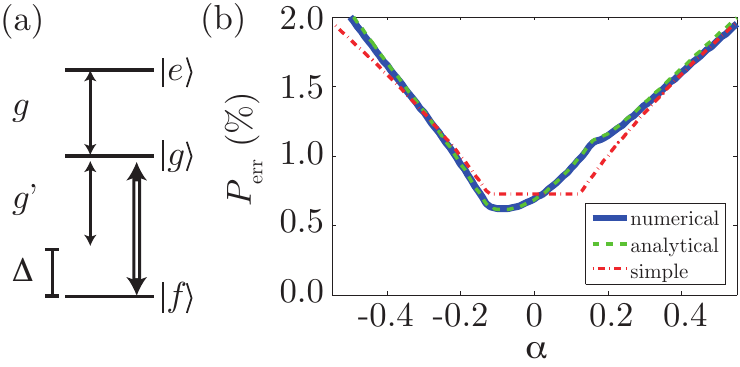}
\caption{Setup. (a) Three level ladder ($\Xi$) system coupled to a cavity field. The cavity field resonantly couples the upper states $\ket{g}$ and $\ket{e}$ of a $\Xi$-system with a coupling constant $g$ (single line), while it is far off resonance by a detuning $\Delta$ from the lower transition between $\ket{f}$ and $\ket{g}$, where it interacts with a coupling constant $g'$. The lower transition is driven by a classical field (double line). %An external control line allows driving the lower transition. %The levels $\ket{g}$ and $\ket{e}$ decay to the level below it with decay rates of $\gamma_1$ and $\gamma_2$ respectively. 
(b) Numerical and analytical results for the error probability for the first step of the protocol as a function of deviation from $\kappa=g$, parameterized by $\alpha=(\kappa-g)/g$. We use the parameters of \cite{rigetti} (coherence time $T_2^*=2/(\gamma_g+\gamma_f)=92\ \mu$s, and anharmonicity $\Delta=(2\pi)206$ MHz), but a reduced coupling constant $g=(2\pi)5.3$ MHz. Full line: numerical results. Dashed line: analytical results. Dashed dotted line: approximate optimum discussed in the text. The pulse length is chosen according to an approximate optimization \cite{supplementary} with $n=8$.
%Equivalent system realized with two qubits coupled to a cavity. Changing to the $\ket{\rm S}$/$\ket{\rm T}$ basis, this system can be mapped to the situation in (a). External control lines allow driving the qubits with opposite phases. This drives the $\ket{00}\leftrightarrow\ket{\rm S}$ transition, but the $\ket{\rm S}\leftrightarrow\ket{11}$ transition is blocked by the coupling of $\ket{11}$ to the cavity.
}
\label{fig:atom_levels}
\end{figure}

We focus on the three lowest levels of the system. 
We assume that the $\Xi$-system is sufficiently anharmonic such that the cavity is resonant with the upper transition between states $\ket{g}$ and $\ket{e}$, which has a coupling constant $g$, but off-resonant from the lower transition between $\ket{f}$ and $\ket{g}$, see Fig. \ref{fig:atom_levels} (a). For now, we  ignore the cavity coupling on the lower transition, %and assume that the cavity only interacts resonantly with the upper transition with a coupling constant $g$, 
deferring the discussion of this off-resonant cavity coupling to later. %Throughout this letter we will assume that the transition frequency of the upper transition is resonant with the cavity frequency, i.e. $\omega_{eg} = \omega_{c}$. 
With this arrangement, we can be in a regime where there is a strong cavity enhancement of the decay of the upper level $\ket{e}$ via the Purcell effect, such that it decays rapidly compared to the two lower levels $\ket{g}$ and $\ket{f}$. %(provided that the coupling constant is much larger than the decay rate of the lower levels and that the cavity decay rate $\kappa$ is chosen accordingly, 
% (see below). 
In essence, this combined system of the $\Xi$-atom and the cavity thus has two metastable levels ($\ket{f}$ and $\ket{g}$) which mimic the stable states of the $\Lambda$-type atom in the proposal of Ref. \cite{transistor}.
To control the system the transition between $\ket{f}$ and $\ket{g}$ is driven by a classical field. Below we describe how these ingredients allow us to implement a single photon transistor. For simplicity we discuss the scheme for a two sided cavity with identical (photon) decay rates $\kappa$ to modes propagating to the left and right. Alternatively, the scheme could be realized with a single sided cavity if additional linear optics elements are used \cite{scattering}.

We describe the system using the ``quantum jump" approach \cite{meystre}, where the evolution is described by  a non-Hermitian Hamiltonian interrupted by quantum jumps at random times. The non-Hermitian Hamiltonian is given by
 \begin{align}
\label{eq:hamiltonian}
H = &\sum_{m=f,g,e}\hbar(\omega_m- i\gamma_m/2)\hat{\sigma}_{mm} +\hbar \omega_c \hat{a}^\dagger \hat a \nonumber\\ + 
&\sum_{m=L,R} \hbar\int dk\, ck \hat{b}_m^\dagger(k)\hat{b}_m(k) + i\hbar g \big[\hat{a}^\dagger \hat{\sigma}^-_{eg} - \hat{a}\hat{\sigma}^+_{eg}\big] \nonumber \\+&\sum_{m=L,R} i\hbar c\sqrt{\kappa/2\pi}\int dk {\left[\hat{b}_m(k) \hat{a}^\dagger - \hat{b}_m^\dagger(k) \hat{a}\right]},
\end{align} 
where the first three terms are the free energy and the last term of the second line is the interaction between the atom and the cavity. The last term is the interaction between the cavity and the external fields, which are represented by operators $\hat{b}_L$ and $\hat{b}_R$ for the field modes on the left and right side of the cavity and $\kappa$ is the total decay rate of the cavity field into the continuum. We assume a linear dispersion relation for the waveguides modes, where $c$ is the velocity of the photons and $\kappa$ is assumed to be frequency independent. The decay rate of the $m$th level $\gamma_m$ represents both relaxation ($T_1$) processes and pure ($T_2$) dephasing  (see supplemental material for  details on the decoherence model \cite{supplementary}).
%Note that $g$ and $\kappa$ can (to some degree) be controlled experimentally by the design of the cavity while $\gamma_1$, and $\gamma_2$ may be more difficult to control. Below we will therefore optimize the cavity decay rate to optimize the performance of the transistor. 

The protocol consists of two steps: first the interaction of the system with the gate photon, and second,  sending in a multi-photon signal field. A key ingredient in our scheme is the difference in the scattering dynamics for photons scattering off the cavity conditioned on the state of the atom. If the atom is in state $|f\rangle$ it does not interact with the cavity field and the incoming resonant photon is transmitted through the cavity. On the other hand, if the atom is in state $|g\rangle$ the atom and the cavity form dressed states at frequencies $\omega_c\pm g$, where $\omega_c$ is the cavity frequency. An incoming photon at the undressed cavity frequency will be off-resonant and will be reflected. The state of the atom thus controls whether the cavity is transmitting the signal or not and thereby acts as a switch. 

\begin{figure}[h]
\centering
\includegraphics{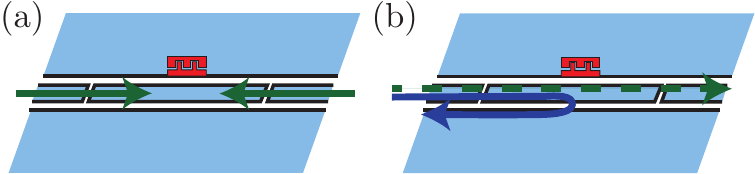}
\caption{Steps of the single-photon transistor operation. (a) First step. A gate field is sent in symmetrically from the left and right. This makes it indistinguishable whether the photon was reflected or transmitted, but still a phase difference is imparted on the atom (see text). This phase difference can be used to flip the atom depending on the presence or absence of a photon in the field. (b) Second step. A strong signal field is sent in from the left. The field is reflected if the atom is in state $\ket{g}$ (full arrow), and transmitted if the atom is in state $\ket{f}$ (dashed arrow). }
\label{fig:protocol}
\end{figure}

We can realize the single photon transistor if we can flip the state of the atom conditioned on the presence of a photon in the first gate pulse. To do this we use the procedure proposed in Ref. \cite{kimble} and realized in Ref. \cite{kim}: We start with the atom initialized in the state $\ket{\Psi_0} = 1/\sqrt{2}(\ket{g} + \ket{f})$ using the classical field on the transition between $\ket{g}$ and $\ket{f}$. Then we send in the gate pulse containing a single photon or not. By taking this gate pulse to be sent in symmetrically as shown in Fig. \ref{fig:protocol} (a) we ensure that we cannot determine the atomic state from whether the pulse has been transmitted or reflected. There is, however, a phase difference between the reflection and transmission such that in a suitable rotating frame the whole evolution is \cite{kimble}
\begin{equation}
\begin{split}
&\ket{0}\ket{\Psi_0} \rightarrow \frac{1}{\sqrt{2}}\ket{0}(\ket{g} + \ket{f}) \\
&\ket{1}\ket{\Psi_0} \rightarrow \frac{1}{\sqrt{2}}\ket{1}(\ket{g} - \ket{f}), \\
\end{split}
\label{eq:idealscat}
\end{equation}
where $\ket{0}$ ($\ket{1}$) denotes a zero (one) photon state of the field. 
Since the phase of the superposition now depends on the absence or presence of a single photon we can use a $\pi/2$-pulse on the $\ket{f} \rightarrow \ket{g}$ transition to map this onto the atomic population. The overall transformation in the first step is then given by
\begin{equation}
\begin{split}
\label{eq:protocol}
&\ket{0}\ket{\Psi_0} \rightarrow \ket{0} \ket{g} \\
&\ket{1}\ket{\Psi_0} \rightarrow \ket{1}\ket{f}, \\
\end{split}
\end{equation}
which exactly realizes a conditional flip of the atomic state.
The second step consists in sending in signal photons %to the system for a time $T'$. In this step the beam of photons is
 %only incident 
from the left as shown in Fig. \ref{fig:protocol} (b). Since the transmission or reflection of these photons is determined by the atomic state, the presence or absence of a photon in the first step controls the transmission in the second step. Hence the whole procedure implements a single photon transistor where the gate %signal 
field controls the propagation of the large signal field for a time only limited by the lifetime $T_1$ of the atomic populations.

%\begin{figure}[h]
%\centering
%\includegraphics[width=0.6\columnwidth]{step2}
%\caption{Second step. A strong field is sent in from the left. The field is reflected if the atom is in state $\ket{g}$ (full arrow), and transmitted if the atom is in state $\ket{f}$ (dashed arrow).}
%\label{fig:step2}
%\end{figure}

In the following we make a detailed analysis of the protocol in the presence of imperfections (extending the analysis of Ref. \cite{kimble} where the protocol was mainly studied numerically). For the first step of the protocol we have a single incoming photon and we can expand the entire wave function on suitable basis states.
% The general state of the system in the case of a single excitations, i.e. with a single photon incoming and the atom in one of the two ground states, is 
%\begin{multline}
%\label{eq:psi}
%\ket{\Psi(t)} = c_a(t)\ket{g,1,0,0} + c_b(t)\ket{e,0,0,0} + c_c(t)\ket{g,0,0,0} + c_d(t)\ket{f,1,0,0} + c_e(t)\ket{f,0,0,0} + \\ \int dk \big[d_k(t) \hat{b}^\dagger_k \ket{g,0,0,0} + d'_k(t) \hat{b}^\dagger_k \ket{f,0,0,0} + 
%e_k(t) \hat{c}^\dagger_k \ket{g,0,0,0} + e'_k(t) \hat{c}^\dagger_k \ket{f,0,0,0}\big],
%\end{multline}
%where the four labels of each ket refer respectively to the state of the atom and to the number of photons in the cavity and in the left and right waveguides.
The Schr\"odinger equation with the Hamiltonian \eqref{eq:hamiltonian} then results in two sets of differential equations for the amplitudes corresponding to the system starting in each of the uncoupled states $\ket{g}$ and $\ket{f}$. 
% the terms in \eqref{eq:psi}, because in the Hamiltonian \eqref{eq:hamiltonian} we have assumed no coupling between $\ket{g}$ and $\ket{f}$. 
From solving the equations we find the reflection coefficient $r(k)$ for a photon with wavenumber $k$ when the atom is in state $\ket{g}$ \cite{supplementary}
\begin{equation}
r(k) = \frac{\delta(k)\big(\delta(k) + i\Gamma\big) - g^2}{\big(\delta(k) + i\Gamma\big)\big(\delta(k) + i\kappa\big) - g^2}.
\label{eq:reflect_coeff_g}
\end{equation}
Here $\delta(k)=ck - \omega_c$ and $\Gamma = (\gamma_e - \gamma_g)/2$. This reflection coefficient is approximately unity in the regime we are interested in $g\gg \sqrt{\kappa\Gamma}, \sqrt{\kappa\delta}, \delta$. Similarly, the transmission amplitude is found to be 
\begin{equation}
\label{eq:transm_coeff_g}
t(k) = \frac{-i\kappa\big(\delta(k) + i\Gamma\big)}{\big(\delta(k) + i\Gamma\big)\big(\delta(k) + i\kappa\big) - g^2},
\end{equation}
which vanishes in the same limit.
When the atom is in state $\ket{f}$ the transmission and reflection coefficient can be found by setting $g=\Gamma=0$ in the above expression 
%\begin{equation}
%r'(k) = \frac{\delta(k)}{\delta(k) + i\kappa} = \frac{-i\delta(k)\kappa + \delta^2(k)}{\kappa^2 +\delta^2(k)}
%\label{eq:reflect_coeff_f}
%\end{equation}
%and
%\begin{equation}
%\label{eq:transm_coeff_f}
%t'(k) = \frac{-i\kappa}{\delta(k) + i\kappa} = \frac{-i\kappa\delta(k) - \kappa^2}{\kappa^2 +\delta^2(k)},
%\end{equation}
and for $\kappa\gg\delta$ these can be then approximated to be $r\approx 0$ and $t\approx -1$. These results justify the evolution in \eqref{eq:idealscat} provided that the bandwidth of the pulse is smaller than ${\rm min}(\kappa,g^2/\kappa)$. [Above we have rescaled all amplitudes by $\exp(\gamma_g t/2)$ and $\exp(\gamma_f t/2)$ to remove the damping of the levels $\ket{g}$ and $\ket{f}$]. 

To get a concrete estimate of the functioning of the transistor we assume the incident gate pulse to have a Gaussian temporal profile
\begin{equation}
\label{eq:gaussian_pulse}
f(t) \propto e^{-(t-T/2)^2/4\sigma_T^2},
\end{equation}
with a width $\sigma_T$.
%and the atomic state is initialized in $\ket{\Psi_0} = 1/\sqrt{2}(\ket{g} + \ket{f})$.
%Since we include the decay of the levels we need to restrict the time of the first protocol. 
We restrict the dynamics to a finite time interval $[0,T]$ and choose the ratio $n=T/\sigma_t$ sufficiently big, e.g., $n=8$, that we can ignore the pulse area outside this interval. %, e.g., we use $n=8$ which excludes a negligible part of the pulse. 
 Assuming equal probabilities to receive a photon or not during the first step we find the average error probability for setting the qubit cavity system
\begin{equation}
\label{eq:err_approx}
P_{{\rm err}} \approx \frac{1}{4}\bigg((\gamma_g + \gamma_f) n \sigma_T + \frac{\kappa(\gamma_e - \gamma_g)}{g^2} + \frac{1}{2\sigma_T^2\kappa^2}\bigg(1 - \frac{\kappa^2}{g^2}\bigg)^2\bigg).
\end{equation}
%where $\gamma_i$ is the decoherence rate of state $\ket{i}$.
This result is valid in the limit where all terms are small. 
It accounts for three sources of error: The first term is given by the decoherence  of the levels $\ket{g}$ and $\ket{f}$. The second term represents the fact that the cavity is not ideal (since this term accounts for decoherence while in state $\ket{e}$ and the full decoherence of level $\ket{g}$ is included in the first term, it is only the additional decoherence $\gamma_e-\gamma_g$ which appears here). The third term is the error due to the finite bandwidth of the pulse which causes an imperfect phase shift with the atom. This term is minimized for $\kappa=g$, which ensures that the dispersion relation for photons at the central frequency is the same regardless of whether the atom is in state $\ket{f}$ or $\ket{g}$. At $\kappa=g$ the outgoing photon wave packets will thus have the same group delay ensuring a maximal overlap, and the error probability only involves a higher order term $1/(\kappa\sigma_T)^6$. With this choice the transistor is efficient for any pulse of temporal width   $1/g\lesssim \sigma_T\lesssim 1/ n(\gamma_g+\gamma_f)$.

So far we have assumed that the coupling between the two lower states can be neglected. 
The detuning $\Delta$ of the cavity from the lower transition equals the anharmonicity of the atom and for large $\Delta$  the coupling  can be taken into account by adding an additional error to the first step
$P_{\rm err}\approx ...+ \frac{1}{4} \tilde{\kappa} n \sigma_T $, where $\tilde{\kappa} = g'^2\kappa/(\Delta^2+\kappa^2)$ and $g'$ is the induced decay rate and coupling constant on the lower transition \cite{supplementary}. 
Experimentally the best coherence properties are currently reached for transmon qubits which have a rather low anharmonicity $\Delta\sim g$ \cite{rigetti,sears,houck}. For typical experimental parameters $g'\sim g$ and the induced decay would be highly detrimental.
It will thus be desirable to reduce the coupling constant (e.g., by placing the atom near a field antinode) to suppress the off-resonant coupling. For a given superconducting system characterized by the anharmonicity $\Delta$ and the  decoherence rates $\{\gamma_i\}$ we can then optimize the performance of the transistor by simultaneously optimizing the coupling constant and the pulse duration. Assuming $\kappa=g$ [and including higher order terms omitted in Eq. (\ref{eq:err_approx})] we find that the optimal error scales as $P_1\sim((\gamma_g+\gamma_f)/\Delta)^{4/7}$. In practice the condition $g=\kappa$ may not be perfectly fulfilled due to fabrication imperfections etc. To characterize this we introduce a small parameter $\alpha=(\kappa-g)/g$, optimize the error probability assuming that we are limited by $\alpha$, and find $P_2\sim\alpha^{2/3}((\gamma_g+\gamma_f)/\Delta)^{4/9}$. The full error will then be $\sim{\rm max}(P_1, P_2)$ (for further details see supplemental material \cite{supplementary}). These expressions show that the error rate can be quite low for realistic systems for which $\Delta\gg \gamma_i$. In particular, recent experiments with 3D confined transmon qubits have shown long coherence times. Taking parameters from Ref. \cite{rigetti} we find that the error rate can be less than $1\%$ for an imbalance $|\alpha|\lesssim 10\%$ (error $1.2\%$ for the parameters of Ref. \cite{sears}). In Fig. \ref{fig:atom_levels} (b) we show the results of a direct numerical simulation of the first step of the protocol along with a detailed analytical theory and the approximate optimization. The results are seen to be in very good agreement. Alternatively, the proposal could also be implemented for transmon systems not confined to 3D cavities. Taking the parameters of Ref. \cite{houck} we find an optimal error of $3.3\%$.

\begin{figure}[htbp]
\begin{center}
 \includegraphics[width=8.5cm]{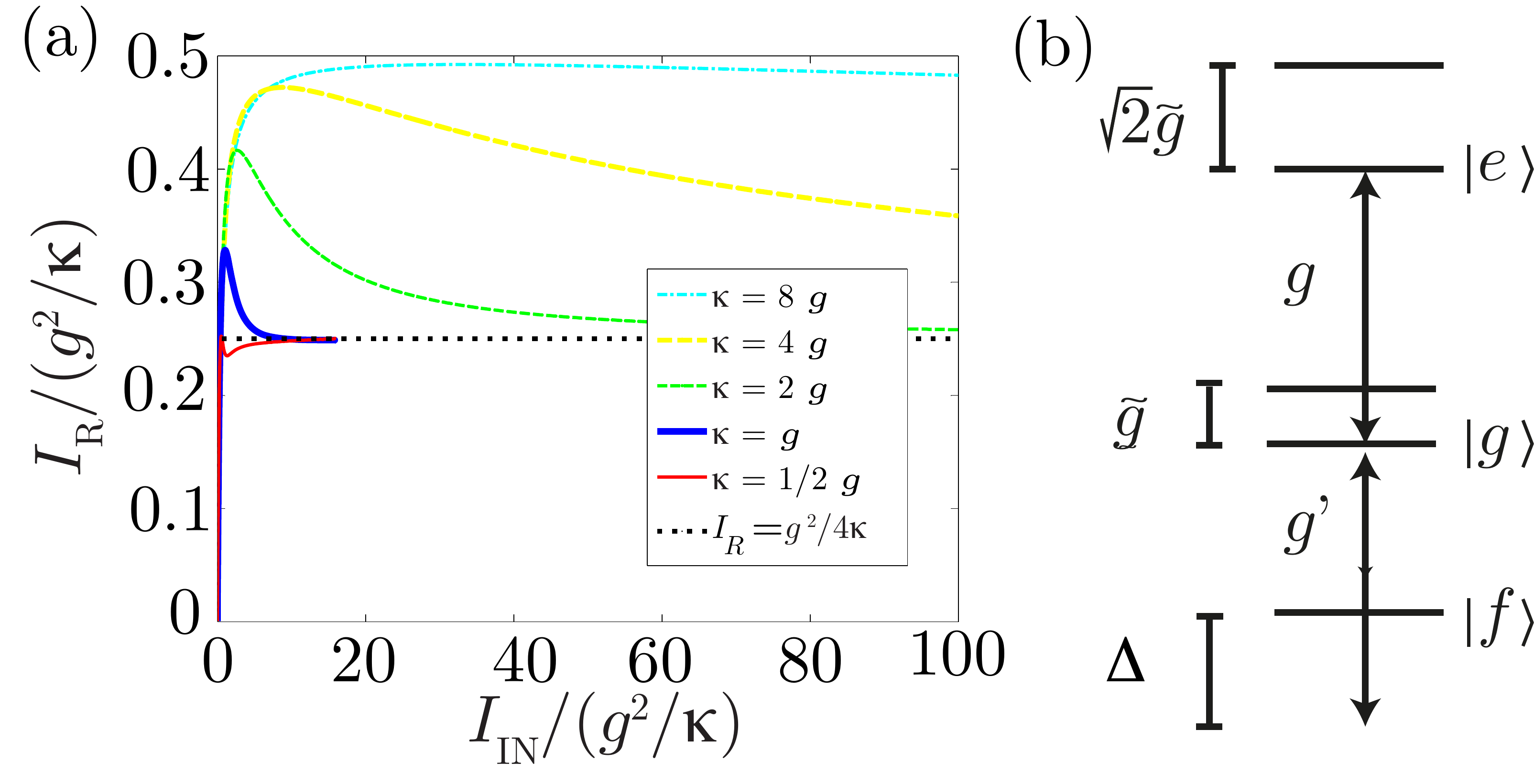}
 % \raisebox{0.9cm}{\includegraphics{effective_anharmonic.pdf}}
\caption{(a) Reflected intensity $I_{{\rm R}}$ for an atom in state $\ket{g}$ as a function of incoming intensity $I_{{\rm IN}}$ obtained from a numerical solution of the master equation for various values of $g/\kappa$. For weak incoming intensities the reflected intensity is equal to the total incoming intensity, but saturates in the range  $g^2/4\kappa$ (dotted line) to $g^2/2\kappa$ as the intensity is increased.  We have assumed $\kappa=g$, and that we only have population decay of the upper level at a rate $\gamma_{e}=1.3\cdot10^{-3} g$ corresponding to the values in Fig. \ref{fig:atom_levels} with $\gamma_e=2\gamma_g$. The saturation of the reflected intensity provides an upper limit to how well one can distinguish an atom being in $\ket{g}$ from an atom being in $\ket{f}$ which has a vanishing reflection. This saturation thus limits the gain of the transistor. (b) Alternative implementation based on a two-level system coupled to  a cavity. Taking the subsystem to have a vacuum Rabi coupling $\tilde{g}$,  the nonlinearity of the Jaynes-Cummings Hamiltonian can mimic our example three-level system.  Denoting the atomic levels by $\tilde{g},\tilde{e}$ and the subsystem cavity Fock states with $\tilde{0}$, etc., we identify $\ket{f} = \ket{\tilde{g},\tilde{0}}, \ket{g} = (\ket{\tilde{g},\tilde{1}} - \ket{\tilde{e},\tilde{0}})/\sqrt{2}, \ket{e} = (\ket{\tilde{g},\tilde{2}} - \ket{\tilde{e},\tilde{1}})/\sqrt{2}$. This yields an anharmonicity parameter $\Delta = 2 \tilde{g} (1 - \sqrt{2})$, while the coupling for the effective three-level system $g$ is determined by the coupling of the anharmonic cavity-subsystem to another cavity and can be as large as desired by appropriate cavity design.}
\label{fig:reflection}
\end{center}
\end{figure}

We now turn to the second step. Here we consider sending in a coherent state of constant amplitude %$\xi$
for a time $T'$. This pulse is resonant with the cavity, but the transmission depends on the atomic state induced by the first signal pulse. The strongest difference in the distribution of the outgoing photons is obtained by sending in a strong beam resonant with the cavity with an intensity $I\gtrsim g^2/\kappa$. If the atom is in state $\ket{f}$ it essentially does not participate in the scattering dynamics and the cavity is just transmitting. If the atom is in state $\ket{g}$ the interaction shifts the cavity out of resonance and the light is reflected. Dealing with this situation in full detail is quite complicated and we therefore limit our calculations to a steady state calculation, which ignores dynamics on a time scale of $1/\kappa \ll T'$ (similar to the calculation in Ref. \cite{nonlinearcavityexp}). We numerically solve the density matrix equation in steady state for various incoming intensities, and evaluate the reflected and transmitted intensities from the steady state solution \cite{supplementary}. With this we find that for $\kappa\sim g$ the maximal reflected  intensity $I_{\rm R}$ saturates in the range $I_{\rm R}=g^2/4\kappa$ to $I_{\rm R}=g^2/2\kappa$ for strong incoming intensities $I_{\rm IN}\gg g^2/\kappa$ and small atomic decay $\gamma_i\ll\kappa$ as show in Fig. \ref{fig:reflection}. The same conclusion is reached from a perturbative calculation valid for strong driving (see supplemental material \cite{supplementary}).
Taking the lower of these two values, the gain of the transistor $G$, defined as the difference in the reflected number of photons condition on an incident photon, is given by $G=T'g^2/4\kappa$. The time $T'$ is limited by the decay of state $\ket{g}$ to $\ket{f}$. %If we want the gain to be the same in all events we should pick a short time $\gamma_{1g} T' \ll 1$ to ensure a small probability of decay. 
The largest average gain is found for $T' \gtrsim T_1$ where we obtain $G=T_1g^2/4\kappa$. % which can be substantial for current experimental setup (note that here it is only the relaxation rate which matters). For current superconducting systems this results in a substantial gain.
Taking the optimized settings derived for the first step of the protocol we find gains of $G \approx 783$, 570, and 116 for the numbers of Refs. \cite{rigetti}, \cite{sears}, and \cite{houck}, respectively. This last step represents the slowest part of the procedure which thus limits the duty cycle of the transistor, but the speed can be increased by reducing $T'$ at the expense of  also reducing the gain.

In summary, we have shown that using superconducting circuit QED it is possible to realize a working photonic transistor sensitive to only a single incoming microwave photon. The device can readily be implemented with existing technology. 
%Taking for instance the 3D transmon setup \cite{rigetti, sears} where recent experiments coherence times $T_{1,2}$ have reached $\approx 30 \mu$s and higher, we find that the error rate of the single-atom realization can be reduced to $7 \%$ for the numbers in Ref. \cite{rigetti} and $\approx 9 \%$ for the numbers in Refs. \cite{sears}. In both setups gains of $G \gtrsim 260$ are possible. For the same numbers, the two-level realization e 
We have focussed on an implementation based on a three level artificial atom but alternatively the procedure could also be implemented in a two-level system in the strong dispersive regime \cite{sears,hatridge} or by using a setup similar to \cite{stannigel} to mimick the three-level atom by a two-level atom strongly coupled to an optical cavity as outlined in Fig. \ref{fig:reflection} (b), opening up our transistor technique to a wide variety of additional systems.
An immediate application of the device proposed here would be the efficient quantum non-demolition detection of individual itinerant microwave photons. In addition, the device could  easily be incorporated into many quantum information protocols working as an efficient processor of information encoded in the photon field. As a particular example, for higher incident photon numbers the device will serve  as a parity detector, detecting the parity of the incident photon number \cite{johanson}. This is an important primitive for several suggested continuous variable quantum information protocols \cite{braunstein,banaszek}.

We thank Michael Foss-Feig for useful comments on the manuscript.
The research leading to these results has received funding from the European Research Council under the European Union's Seventh Framework Programme (FP/2007-2013) / ERC Grant Agreement n. 306576, from the Studienstiftung des deutschen Volkes, and from the international PhD-fellowship program "la Caixa-Severo Ochoa". JMT thanks E. Waks, M. Hafezi, and P. Kwiat for illuminating conversations, and the kind hospitality of the NBI during his visit. Funding is provided by the NSF through the Physics Frontier Center at the JQI and U.S. Army Research Office Multidisciplinary University Research Initiative award W911NF0910406.

%We remark that our three-level atomic model can be realized with a subsystem comprising a two-level atom strongly coupled to an optical cavity, opening up our transistor technique to a wide variety of additional systems. Specifically, taking the subsystem to have a vacuum Rabi coupling $\tilde{g}$, we use the nonlinearity of the Jaynes-Cummings hamiltonian to mimic our example three-level system.  Denoting the atomic levels by $\tilde{g},\tilde{e}$ and the subsystem cavity Fock states with $\tilde{0}$, etc., we identify $\ket{f} = \ket{\tilde{g},\tilde{0}}, \ket{g} = (\ket{\tilde{g},\tilde{1}} - \ket{\tilde{e},\tilde{0}})/\sqrt{2}, \ket{e} = (\ket{\tilde{g},\tilde{2}} - \ket{\tilde{e},\tilde{1}})/\sqrt{2}$. This yields an anharmonicity parameter $\Delta = 2 \tilde{g} (1 - \sqrt{2})$, while the coupling for the effective three-level system $g' = g / \sqrt{2}$ is determined by the cavity-subsystem cavity coupling rate and can be as large as desired by appropriate cavity design.  However, in practice using this subsystem approach requires stringent constraints on the subsystem atom-cavity coupling and damping: even stronger coupling is required than for the true three-level atom case.

\end{document}

\title{A single photon transistor based on superconducting systems. Supplementary Information}
\author{Marco T. Manzoni}
\affiliation{QUANTOP -- Danish Quantum Optics Center, The Niels Bohr Institute, University of Copenhagen, Blegdamsvej 17, DK-2100 Copenhagen \O, Denmark}
\affiliation{ICFO-Institut de Ciencies Fotoniques, 08860 Castelldefels (Barcelona), Spain} 
\author{Florentin Reiter}
\affiliation{QUANTOP -- Danish Quantum Optics Center, The Niels Bohr Institute, University of Copenhagen, Blegdamsvej 17, DK-2100 Copenhagen \O, Denmark}
\author{Jacob Taylor}
\affiliation{Joint Quantum Institute/National Institute of Standards and Technology, 100 Bureau Dr MS 8410, Gaithersburg, MD  20899}
\author{Anders S. S\o rensen}
\affiliation{QUANTOP -- Danish Quantum Optics Center, The Niels Bohr Institute, University of Copenhagen, Blegdamsvej 17, DK-2100 Copenhagen \O, Denmark}
%\title{Supplemental Information}
\date{\today} % Activate to display a given date or no date (if empty),
% otherwise the current date is printed 

\maketitle

Below equation references such as ``(1)'' will refer to equations from the main text, whereas equations labelled, e.g., ``(S1)'' will refer to equations in this supplementary material. 

\section{Details of the description of decoherence}
In this section we will give details of the decoherence model assumed in the main paper and discuss how the parameters of our model can be related to the decay of various density matrix elements.
In general the decay of the density matrix is described by the master equation. Here we for simplicity restrict ourselves to a Markovian master equation, in which case the decoherence is characterized by five decoherence rates, two for the decay of the populations $\rho_{ee}$ and $\rho_{gg}$ and three for the decay of the coherences $\rho_{eg}$, $\rho_{ef}$, and $\rho_{gf}$. We call the former $\gamma_{1e}$ and $\gamma_{1g}$, while we write the latter as $(\gamma_{1i}+\gamma_{1j})/2 + \tilde\gamma_{ij}/2$ (with $\gamma_{1f} = 0$ for the ground state). Here $\tilde\gamma_{ij}/2$ represents a dephasing of the off-diagonal elements of the density matrix between states $\ket{i}$ and $\ket{j}$ in addition to the decay induced by the decay of the populations.

We introduce the Lindblad operators $\hat{c}_{2m}=\sqrt{\gamma_{2m}}\sigma_{mm}$ to describe the dephasing, and $\hat{c}_{1e}=\sqrt{\gamma_{1e}}\sigma^-_{eg}$ and $\hat{c}_{1g}=\sqrt{\gamma_{1g}}\sigma^-_{gf}$ which describe the decay. In writing the last Lindblad operators we have assumed that the decay only takes the system from a state to the state immediately below it. With these decoherence mechanisms we get the following equations for the six independent elements of the density matrix:
\begin{equation}
\label{eq:master_equation}
\begin{split}
&\dot{\rho}_{ee}(t) = -\frac{i}{\hbar}[H_s, \rho_{ee}(t)] - \gamma_{e1}\rho_{ee}(t) \\
&\dot{\rho}_{gg}(t) = -\frac{i}{\hbar}[H_s, \rho_{gg}(t)] - \gamma_{g1}\rho_{gg}(t) + \gamma_{e1}\rho_{ee}(t)\\
&\dot{\rho}_{ff}(t) = -\frac{i}{\hbar}[H_s, \rho_{ff}(t)] + \gamma_{g1}\rho_{gg}(t) \\
&\dot{\rho}_{eg}(t) = -\frac{i}{\hbar}[H_s, \rho_{eg}(t)] - \frac{\gamma_{e1} + \gamma_{g1} + \gamma_{e2} + \gamma_{g2}}{2}\,\rho_{eg}(t) \\
&\dot{\rho}_{gf}(t) = -\frac{i}{\hbar}[H_s, \rho_{gf}(t)] - \frac{\gamma_{g1} + \gamma_{g2} + \gamma_{f2}}{2}\,\rho_{gf}(t) \\
&\dot{\rho}_{ef}(t) = -\frac{i}{\hbar}[H_s, \rho_{ef}(t)] - \frac{\gamma_{e1} + \gamma_{e2} + \gamma_{f2}}{2}\,\rho_{ef}(t), \\
\end{split}
\end{equation} 
From here we can find the relation between the decay rates of our model and the dephasing induced decay of the coherence
\begin{equation}
\begin{split}
&\tilde \gamma_{eg} = \gamma_{e2} + \gamma_{g2}, \\
&\tilde \gamma_{gf} = \gamma_{g2} + \gamma_{f2}, \\
&\tilde \gamma_{ef} = \gamma_{e2} + \gamma_{f2}, \\
\end{split}
\end{equation}
which can be inverted to give
\begin{equation}
\begin{split}
&\gamma_{e2} = (\tilde\gamma_{eg} + \tilde\gamma_{ef} - \tilde\gamma_{gf})/2,\\
&\gamma_{g2} = (\tilde\gamma_{eg} + \tilde\gamma_{gf} - \tilde\gamma_{ef})/2,\\
&\gamma_{f2} = (\tilde\gamma_{ef} + \tilde\gamma_{gf} - \tilde\gamma_{eg})/2.
\end{split}
\end{equation} 

In the quantum jump picture we get the effective non-Hermitian Hamiltonian by adding the terms $-i\hbar\hat{c}^\dagger_j\hat{c}_j/2$, where $\hat{c}_j$ are the Lindblad operators, to the system Hamiltonian \cite{meystre}. In
our case this corresponds to adding the terms $i\hbar(\gamma_{1m} + \gamma_{2m})/2\,\sigma_{mm}$ with $m = e,g,f$ to the atom-cavity Hamiltonian to get the full no-jump Hamiltonian \eqref{eq:hamiltonian}.

In the first step of the protocol we first prepare a superposition $(\ket{g}+\ket{f})/\sqrt{2}$ by a $\pi/2$-pulse and then later apply a second $\pi/2$-pulse. To obtain the correct behavior of the transistor it is essential that we maintain the correct phase relation between $\ket{g}$ and $\ket{f}$. As a consequence the quantity appearing in our formulas is the sum of the full decoherence rates $\gamma_g+\gamma_f$ which can be directly related to the decay of the off-diagonal elements of the density matrix, c.f. Eq. \eqref{eq:master_equation}. For a coherence decaying as $\rho_{fg}(t)=\rho_{fg}(t=0)\exp(-t/T_2^*)$ we have $\gamma_g+\gamma_f=2/T_ 2^*$. For the second step on the other hand, it is the population decay which is the most important parameter. 

Below and in the main text we use values for the decoherence parameters, e.g. $T_2^*$, extracted from real experimental setups. In reality the experiments may, however, not exhibit the Markovian noise that we assume here, but may rather be of the type $\sim\exp(-(t/T_2^*)^2)$. Since $\exp(-(t/T_2^*)^2)$ approaches unity much more rapidly than $\exp(-t/T_2^*)$ for $t\ll T_2^*$, the Markovian assumption that we make is actually a worst case scenario. For a different decoherence model we would therefore find a better performance for the same parameters. 

\section{Single photon dynamics}

In this section we consider the dynamics of the first step, in which a single photon is incident to the cavity from either the left or the right and interacts with the atom which is prepared in one of the two lower states. The linearity of the Schr\"odinger equation permits us to treat these different cases separately and then to combine the results to describe the superposition of the initial states.

We notice that the number of excitations operator 
\begin{equation}
\hat{N} = \sigma_{ee} + \hat{a}^\dagger\hat{a} + \sum_{m=L,R} \hbar\int dk\, \hat{b}_m^\dagger(k)\hat{b}_m(k) 
\end{equation}
commutes with the effective Hamiltonian \eqref{eq:hamiltonian}. This allows us to restrict our treatment to the subspace of the overall Hilbert space containing either zero or one excitation. The generic form of the wavefunction in this case is
\begin{multline}
\label{eq:psi}
\ket{\Psi(t)} = c_a(t)\hat{a}^\dagger \ket{g,\O} + c_b(t)\ket{e,\O} + c_c(t)\ket{g,\O} + c_d(t)\hat{a}^\dagger\ket{f,\O} + c_e(t)\ket{f,\O} + \\ \int dk \big[d_k(t) \hat{b}^\dagger_L(k) \ket{g,\O} + d'_k(t) \hat{b}^\dagger_L(k) \ket{f,\O} + 
e_k(t) \hat{b}^\dagger_R(k) \ket{g,\O} + e'_k(t) \hat{b}^\dagger_R(k) \ket{f,\O}\big],
\end{multline}
where we have simplified the notation by introducing $\ket{i,\O}$ to denote the states with the atom in state $i=e,f,g$ and all field modes in the vacuum.
%where the four labels of each ket refer respectively to the state of the atom and to the number of photons in the cavity and in the left and right waveguides.

If the atom starts out in state $\ket{g}$ and the gate photon is coming from the left at time $t = 0$, all the coefficients except $d_k$ vanish.
% the following comment is not relevant yet as we look at scattering in general:, and $d_k(t=0)$ is the Fourier transform of Eq. [REF. TO EQ.8 OF THE MAIN TEXT]. 
If the atom starts out in $\ket{f}$ the non-vanishing coefficients are $d'_k$. % which are then given by the same expressions as for $d_k$ in the previous case. 
With the ansatz above the Schr\"odinger equation results in a set of coupled differential equations. With the atom initially in state $\ket{g}$ we get
\begin{equation}
\begin{split}
\label{eq:system_1}
&\dot{c}_a = -i(\omega_g + \omega_c - i\gamma_g/2)c_a + g c_b + c\sqrt{\kappa/2\pi}\int dk\,d_k + c\sqrt{\kappa/2\pi}\int dk\,e_k \\
&\dot{c}_b = -i(\omega_e - i\gamma_e/2)c_b - g c_a \\
&\dot{d}_k = -\sqrt{\kappa/2\pi} c_a - i(\omega_g + ck - i\gamma_g/2)d_k \\
&\dot{e}_k = -\sqrt{\kappa/2\pi} c_a - i(\omega_g + ck - i\gamma_g/2)e_k. \\
\end{split}
\end{equation}
We assume that the cavity is resonant with the upper atomic transition, i.e. $\omega_{eg} = \omega_e - \omega_g = \omega_c$ and move to a kind of interaction picture, where we replace the amplitudes $c_{a,b}(t)$, $d_k(t)$, and $e_k(t)$ in Eq. \eqref{eq:system_1} by slowly varying amplitudes $C_{a,b}(t)=c_{a,b}(t)e^{i(\omega_e - i\gamma_g/2)t}$, etc. Note that we also rescale the amplitudes of state $\ket{e}$ by the decay rate for state $\ket{g}$.
In this way we get the following equations for the new amplitudes
\begin{equation}
\label{eq:sistema_3}
\begin{split}
&\dot{C}_a = g \,C_b + c\sqrt{\kappa/2\pi}\int dk\,D_k + c\sqrt{\kappa/2\pi}\int dk\,E_k \\
&\dot{C}_b = -(\gamma_e - \gamma_g)/2\, C_b - g C_a \\
&\dot{D}_k = -\sqrt{\kappa/2\pi} C_a - i(ck - \omega_c)D_k \\
&\dot{E}_k = -\sqrt{\kappa/2\pi} C_a - i(ck - \omega_c)E_k. \\
\end{split}
\end{equation} 
with the initial conditions $C_a(t = 0) = C_b(t = 0) = C_c(t = 0)=0$. We formally integrate the equation for $D_k$ and find
\begin{equation}
\label{eq:d_k}
D_k(t) = e^{-i(ck-\omega_c)t}D_k(0) -\sqrt{\kappa/2\pi}\int^t_0 dt'\,e^{-i(ck-\omega_c)(t-t')}C_a(t') 
\end{equation}
and similarly for $E_k$. The equation for $C_a$ can also be formally integrated to give
\begin{equation}
\label{eq:c_a}
C_a(t) = g\int^t_{0} dt'\,C_b(t') + c\sqrt{\kappa/2\pi}\int dk\,\int^t_0 dt'\,(D_k(t') + E_k(t'))
\end{equation}
If we insert \eqref{eq:d_k} in \eqref{eq:c_a} we get 
\begin{multline}
\label{eq:c_a2}
C_a(t) = g\int^t_{0} dt'\,C_b(t') + \sqrt{\kappa}\int^t_{0} dt'\,\big(D^{IN}(t') + E^{IN}(t')\big) - \\
- \frac{\kappa c}{\pi}\int dk\,\int^t_{0} dt'\,\int^{t'}_{0} dt''\,e^{-i(ck-\omega_c)(t'-t'')}C_a(t''),
\end{multline}
where we have defined 
\begin{equation}
\label{eq:in_field}
D^{IN}(t) = \frac{c}{\sqrt{2\pi}}\int dk\,e^{-i(ck-\omega_c)t} D_k(0),
\end{equation}
which is the (rescaled) photonic incoming field amplitude at the left entrance of the cavity. Similarly $E^{IN}(t)$ is the field at the right entrance.
In the last integral in Eq. \eqref{eq:c_a2} we use the Markov approximation and get
\begin{equation}
\label{eq:c_a3}
C_a(t) = g\int^t_{0} dt'\,C_b(t') + \sqrt{\kappa}\int^t_{0} dt'\,\big(D^{IN}(t') + E^{IN}(t')\big) - \kappa \int^t_{0} dt'\,C_a(t'),
\end{equation}
which means that the only unknown in the equation for $C_a$ is $C_b$, since both $D^{IN}(t)$ and $E^{IN}(t)$ are known from the initial conditions. Taking the time derivative of \eqref{eq:c_a3} and the expression for $\dot{C}_b$ \eqref{eq:sistema_3} we get a system of two first order coupled differential equations for $C_a$ and $C_b$
\begin{equation}
\begin{split}
\label{eq:system_2}
&\dot{C}_a = -\kappa\, C_a(t) + g\, C_b(t) + \sqrt{\kappa}\big(D^{IN}(t) + E^{IN}(t)\big) \\
&\dot{C}_b = -(\gamma_e-\gamma_g)/2\, C_b(t) - g \,C_a(t), \\
\end{split}
\end{equation}
where $D^{IN}$ and $E^{IN}$ are the temporal shapes of the external photonic fields in the left and right waveguide. Fourier transforming and using the boundary condition 
\begin{equation}
\label{eq:boundary}
D^{OUT}(t) - D^{IN}(t) = -\sqrt{\kappa} C_a(t),
\end{equation}
for the left waveguide, and a similar one for the right waveguide, we obtain the reflection and transmission coefficients in Eqs. \eqref{eq:reflect_coeff_g} and \eqref{eq:transm_coeff_g}.

If the atom is initially in $\ket{f}$, the 
Schr\"odinger equation gives the differential equations
\begin{equation}
\begin{split}
\label{eq:system_2}
&\dot{c}_d = -i(\omega_c - i\gamma_f/2)\, c_d + c\sqrt{\kappa/2\pi}\int dk\,d_k' + c\sqrt{\kappa/2\pi}\int dk\,e_k' \\
&\dot{d}'_k = -i(ck - i\gamma_f/2)\,d_k' - \sqrt{\kappa/2\pi} c_d \\
&\dot{e}'_k = -i(ck - i\gamma_f/2)\,e_k' - \sqrt{\kappa/2\pi} c_d.
\end{split}
\end{equation}
If we manipulate these expressions in the same way as above we get the coefficients
\begin{equation}
r'(k) = \frac{\delta(k)}{\delta(k) + i\kappa} = \frac{-i\delta(k)\kappa + \delta^2(k)}{\kappa^2 +\delta^2(k)}
\label{eq:reflect_coeff_f}
\end{equation}
and
\begin{equation}
\label{eq:transm_coeff_f}
t'(k) = \frac{-i\kappa}{\delta(k) + i\kappa} = \frac{-i\kappa\delta(k) - \kappa^2}{\kappa^2 +\delta^2(k)},
\end{equation}
which is the same as the results above with $g = \Gamma = 0$.

\section{Error probability}

Here we will evaluate the error in the first step of the protocol. In essence the calculation of the error requires evaluating the probability that the atom ends up in state $\ket{g}$ in the presence of the control photon as well as the probability that it ends up in state $\ket{f}$ in the absence of the photon. % In the main paper we have showed how to determine the error probability from a calculation of the deterministic no-jump evolution and 
Below we shall evaluate these probabilities. For this purpose we look in detail at the sequence of operations of the first step, using the results of the previous section.
%of this Supplemental Information for the dynamics generated by the interaction of a single photon 
%with the atom-cavity system.

We first consider that a photon is incident during the first step of the protocol. For $t < 0$ we then have a photon pulse consisting of a symmetric superposition of a pulse propagating towards the cavity from the left and from the right. Assuming that the atom is in the lower state $\ket{f}$ before the pulse the initial state is
\begin{equation}
\ket{\psi}^{(1)}_{t=0^-} = \int dk\, d_k(t=0^-)\big(\hat{b}^{\dagger}_L(k) +\hat{b}^{\dagger}_R(k)\big) \ket{f,\O}.
\end{equation}
At $t = 0$ the classical pulse mixes the atomic states so that the state becomes
\begin{multline}
\ket{\psi}^{(1)}_{t=0^-} \xrightarrow{\pi/2} \ket{\psi}^{(1)}_{t=0} = \frac{1}{\sqrt{2}}\int dk\,d_k(t=0)\big(\hat{b}^{\dagger}_L(k) + \hat{b}^{\dagger}_R(k))\big(\ket{g,\O} + \ket{f,\O}\big). 
%= \\
%= \frac{1}{\sqrt{2}}\int dk\,\bigg(d_k(t=0)\big(\hat{b}^{\dagger}_L(k) + \hat{b}^{\dagger}_R(k))\ket{g,\O} + d'_k(t=0)\big(\hat{b}^{\dagger}_L(k) + \hat{b}^{\dagger}_R(k))\ket{f,\O}\bigg),
\end{multline}
Hence the initial condition is specific by $d'_k(t=0)=d_k(t=0)$ with $d_k(t=0)$ given by the Fourier transform of Eq. \eqref{eq:gaussian_pulse}. 
%where in the last line we have changed the name to the amplitude relative to the atomic state $\ket{f}$.
After the pulse the state evolves for a time $T$ and the second classical pulse performs the transformation
\begin{equation}
\begin{split}
&\ket{g} \xrightarrow{\pi/2} \frac{1}{\sqrt{2}}(\ket{g} - \ket{f}) \\
&\ket{f} \xrightarrow{\pi/2} \frac{1}{\sqrt{2}}(\ket{g} + \ket{f}). \\
\end{split}
\end{equation}
The final state is then 
\begin{multline}
\label{eq:final_state_1}
\ket{\psi}^{(1)}_{t=T} = \frac{1}{2}\int dk \bigg(d_k(T)(\hat{b}^{\dagger}_L(k) + \hat{b}^{\dagger}_R(k))(\ket{g,\O} - \ket{f,\O}) + c_a(T)\hat{a}^\dagger(\ket{g,\O} - \ket{f,\O}) + \\
+ \sqrt{2}c_b(T)\ket{e,\O} + d_k'(T)(\hat{b}^{\dagger}_L(k) + \hat{b}^{\dagger}_R(k))(\ket{g,\O} + \ket{f,\O}) + c_d(T)\hat{a}^\dagger(\ket{g,\O} + \ket{f,\O})\bigg].
\end{multline}

 %THE NEXT STUFF WE HAVE EXPLAINED IN THE MAIN TECT SO WE DON'T NEED TO REPEAT (ALTHOUGH WE SHOULD CONSIDER TAKING IT OUT OF THE MAIN PAPER AND PUTTING IT HERE):%This expression includes the fact that we have so far only considered the non-Hermitian evolution, so that at time $T$ the sum of the probabilities for the atomic states are less than unity $p < 1$. Since the last $\pi/2$-pulse
%at time $T$ mixes the two atomic states we can find the normalized probabilities as $P_i = P'_i + (1 - p)/2$.
The probabilities $P'_g$ and $P'_f$ to be in states $\ket{g}$ and $\ket{f}$ are readily obtained from the expression for the final state \eqref{eq:final_state_1}
\begin{equation}
\label{eq:P'_g}
P'_g = \frac{1}{2}\int dk \bigg[2|d_k(T) + d_k'(T)|^2 + |c_a(T) + c_d(T)|^2\bigg]
\end{equation}
and
\begin{equation}
\label{eq:P'_f}
P'_f = \frac{1}{2}\int dk \bigg[2|d_k(T) - d_k'(T)|^2 + |c_d(T) - c_a(T)|^2\bigg].
\end{equation}
The last two contributions in Eqs. \eqref{eq:P'_g} and \eqref{eq:P'_f} originate from the fact at time $T$ the excitation could be still be present as a cavity excitation. We find that the amplitudes $c_a$ and $c_d$ are proportional
to $e^{-(t-T/2)^2/4\sigma_T^2}$ so that for a sufficiently large ratio $n = T/\sigma_T$ these contributions are completely negligible, and we shall ignore them. Similarly we also ignore the probability that the atom is in state $\ket{e}$ at time $t=T$ since this is suppressed by a similar factor.

To find the amplitudes $d_k(T)$ and $d'_k(T)$ we use
\begin{equation}
\label{eq:final_fields_2}
\begin{split}
&d_k(T) = \big(r(k) + t(k)\big)d_k(0)e^{-\gamma_g T/2} \\
&d'_k(T) = \big(r'(k) + t'(k)\big)d'_k(0)e^{-\gamma_f T/2} \\
\end{split}
\end{equation}
with the explicit form of the coefficients given in the paper. With this we have the probabilities as an integral over k, but we cannot directly solve the integral as the coefficients are quite involved. Instead we shall do a perturbative expansion of the expression which allows us to extract the main physical effects. Thus we need to consider $t(k) + r(k)$. We expand the denominator and we arrive at 
\begin{equation}
\label{eq:d_k_app}
t(k) + r(k) \approx 1 - \frac{2\kappa\Gamma}{g^2} + \frac{2i\kappa\delta(k)}{g^2} - \frac{2\kappa^2\delta(k)^2}{g^4},
\end{equation}
where we have omitted some other terms which are much smaller than the ones kept under the assumptions used in the main text, i.e. $g, \kappa \gg \gamma_m, \delta$. 
Using similar arguments we can expand $t'(k) + r'(k)$ to get 
\begin{equation}
\label{eq:d'_k_app}
t'(k) + r'(k) \approx -1 - \frac{2i\delta(k)}{\kappa} + \frac{2\delta(k)^2}{\kappa^2}.
\end{equation}
We now insert Eqs. \eqref{eq:final_fields_2} with \eqref{eq:d_k_app} and \eqref{eq:d'_k_app} in Eqs. \eqref{eq:P'_g} and \eqref{eq:P'_f} and find that
\begin{equation}
P'_g = \frac{1}{4}\int dk f_0(k)\bigg|e^{-\gamma_g T/2}\bigg(1 - \frac{2\kappa\Gamma}{g^2} + \frac{2i\kappa\delta(k)}{g^2} - \frac{2\kappa^2\delta(k)^2}{g^4}\bigg) + e^{-\gamma_f T/2}\bigg(-1 - \frac{2i\delta(k)}{\kappa} + \frac{2\delta(k)^2}{\kappa^2}\bigg)\bigg|^2
\end{equation}
and
\begin{equation}
P'_f = \frac{1}{4}\int dk f_0(k)\bigg|e^{-\gamma_g T/2}\bigg(1 - \frac{2\kappa\Gamma}{g^2} + \frac{2i\kappa\delta(k)}{g^2} - \frac{2\kappa^2\delta(k)^2}{g^4}\bigg) - e^{-\gamma_f T/2}\bigg(-1 - \frac{2i\delta(k)}{\kappa} + \frac{2\delta(k)^2}{\kappa^2}\bigg)\bigg|^2,
\end{equation}
where $f_0(k)$ is a normalized gaussian with width $1/2\sigma_T$.

The above expressions only give the no-jump evolution and to find the true probabilities we have to add the quantum jumps which occur with a probability $1-P_f'-P_g'$. Independent of which jumps occur we always end up in an incoherent mixture of $\ket{g}$ and $\ket{f}$. After the last $\pi/2$-pulse these populations will be equally distributed between $\ket{g}$ and $\ket{f}$ and we can thus find the total probabilities by 
\begin{equation}
P_g = P'_g + \frac{1}{2}(1 - P'_g - P'_f) = \frac{1}{2}(1 + P'_g - P'_f).
\end{equation}
when we include the no-jump evolution.

Inserting the expressions for $P'_g$ and $P'_f$ in the formula for $P_g$ we finally find the probability of error in the case that the control photon is present
\begin{equation}
P^{(1)}_{err} = \frac{1}{2}\bigg[1 - e^{-(\gamma_g + \gamma_f) T/2}\bigg(1 - \frac{2\kappa\Gamma}{g^2} - \frac{\kappa^2}{2g^4\sigma_T^2} - \frac{1}{2\kappa^2\sigma_T^2} + \frac{1}{\sigma_T^2 g^2} \bigg)\bigg].
\end{equation}

Let us now consider the case in which the control photon is not present. In this case the evolution is much simpler. At time $t = 0$ the system is prepared in the state
\begin{equation}
\ket{\psi}^{(0)}_{t=0} = \frac{1}{\sqrt{2}}\big(\ket{g,\O} + \ket{f,\O}\big).
\end{equation}
The time evolution in this case simply consists in the decay of the amplitude of the two states. At time $t = T$, immediately before the second $\pi/2$-pulse, the state of the system is
\begin{equation}
\ket{\psi}^{(0)}_{t=T} = \frac{1}{\sqrt{2}}\big(e^{-\gamma_g T/2}\ket{g,\O} + e^{-\gamma_f T/2}\ket{f,\O}\big),
\end{equation}
and after the classical pulse we get
\begin{equation}
\label{eq:final_sta_0}
\ket{\psi}^{(0)}_{FIN} = \frac{1}{2}\big(e^{-\gamma_g T/2}(\ket{g,\O} - \ket{f,\O}) + e^{-\gamma_f T/2}(\ket{g,\O} + \ket{f,\O})\big).
\end{equation}
From Eq. \eqref{eq:final_sta_0} we then get the probabilities for the atomic states
\begin{equation}
P'_g = \frac{1}{4}\big|e^{-\gamma_g T/2} + e^{-\gamma_f T/2}\big|^2
\end{equation}
and 
\begin{equation}
P'_f = \frac{1}{4}\big|e^{-\gamma_g T/2} - e^{-\gamma_f T/2}\big|^2.
\end{equation}
Including the jump evolution, the probability of error for the case of no control photon is thus
\begin{equation}
P^{(0)}_{err} = (1 + P'_f - P'_g)/2=
\frac{1}{2}\big(1 - e^{-(\gamma_g + \gamma_f)T/2}\big).
\end{equation}

Assuming that the transistor is used in a way such that the probabilities to have or not to have a control field are equal, the average error probability of the first step is 
\begin{multline}
\label{eq:error}
P_{err} = \frac{1}{2}\big(P^{(0)}_{err} + P^{(1)}_{err}\big) = \\
= \frac{1}{4}\bigg[2 - e^{-(\gamma_g + \gamma_f)n\sigma_T/2}\bigg(\big(2 - \frac{2\kappa\Gamma}{g^2} - \frac{\kappa^2}{2g^4\sigma_T^2} - \frac{1}{2\kappa^2\sigma_T^2} + \frac{1}{\sigma_T^2 g^2} \bigg)\bigg].
\end{multline}
If we expand the exponential to the first order we can rewrite Eq. \eqref{eq:error} as 
\begin{equation}
\label{eq:err_approx}
P_{err} \approx \frac{1}{4}\bigg((\gamma_g + \gamma_f)n\sigma_T + \frac{2\kappa\Gamma}{g^2} + \frac{1}{2\kappa^2\sigma_T^2} + \frac{\kappa^2}{2g^4\sigma_T^2} - \frac{1}{\sigma_T^2 g^2} \bigg).
\end{equation}
The last two terms can be written as 
\begin{equation}
 \frac{1}{2\kappa^2\sigma_T^2} + \frac{\kappa^2}{2g^4\sigma_T^2} - \frac{1}{\sigma_T^2 g^2} = \frac{1}{2\sigma_T^2\kappa^2}\bigg(1 - \frac{\kappa^2}{g^2}\bigg)^2
\end{equation}
which is the formula given in the paper.

\section{Effective Hamiltonian for the coupling of the two lower states}

In the paper and the previous sections of this Supplemental Information we assumed that the states $\ket{g}$ and $\ket{f}$ are completely decoupled so that in the quantum jump Hamiltonian \eqref{eq:hamiltonian} no term which couples these states is present. The reason for the decoupling relies on the fact that the atom is assumed to have a very large anharmonicity, such that the transition frequency $\omega_{gf}$ is far detuned from the cavity frequency $\omega_c = \omega_{eg}$. 

A more general description of the system would include a second Jaynes-Cummings-like term in the Hamiltonian to provide a coupling between the state $\ket{g,0}$ and the state $\ket{f,1}$, where the first labels refer to the atomic state and the second to the number of photons in the cavity. The energies of these states are respectively $\omega_g$ and $\omega_f + \omega_c - i\kappa/2$ where, keeping the quantum jump approach, the imaginary part accounts for the decay of the states to the lower state $\ket{f,0}$. The detuning of the transition is given by the difference of the energies and is $\tilde{\Delta} = \omega_{gf} - \omega_c + i\kappa/2 = \Delta + i\kappa/2$ where we $\Delta$ is the real part. The inclusion of such a coupling considerably complicates the dynamics, because the state $\ket{g}$ can now decay by emitting a cavity photon, enlarging the Hilbert space with the introduction of two-cavity photons states, increasing the difficulty of solving the equations and more importantly makes the definitions of transmission and reflection coefficients more problematic.

In the case of a far detuned but dipole allowed transition, as is our case, we can use an effective Jaynes-Cummings Hamiltonian for the far off-resonant interactions \cite{knight}. It can be shown that if $\Delta \gg g_{gf}$ the Jaynes-Cummings interaction Hamiltonian for large detuning can be approximated by
\begin{equation}
\label{eq:hamiltonian_dc}
H_{gf}^{eff} = \frac{\hbar g_{gf}^2}{\tilde{\Delta}}\big[\sigma_{gg} + \hat{a}^\dagger\hat{a}(\sigma_{gg} - \sigma_{ff})\big] = \\
\hbar(\tilde{\chi} - i\tilde{\kappa}/2)\big[\sigma_{gg} + \hat{a}^\dagger\hat{a}(\sigma_{gg} - \sigma_{ff})\big],
\end{equation}
where $g_{gf}$ is the coupling constant of the transition $\ket{g} \leftrightarrow \ket{f}$, and we have defined 
\begin{equation}
\tilde{\chi} = \frac{g_{gf}^2\Delta}{\Delta^2 + \kappa^2}
\end{equation}
and
\begin{equation}
\tilde{\kappa} = \frac{2g_{gf}^2\kappa}{\Delta^2 + \kappa^2}.
\end{equation}
We now turn to the reflection and transmission coefficients for the case where the atom is in $\ket{g}$. Here, we replace the former resonance condition $\omega_c = \omega_{eg}$ by the condition $\omega_c = \omega_e - \omega_g - \tilde{\chi}$. Following the above calculations with the inclusion of the new terms it can be shown that the coefficients of reflection and transmission become
\begin{equation}
\label{eq:transmission_g_dc}
\tilde{t}(k) = \frac{\delta(k)^2 + \tilde{\chi}(\delta(k) + i(\Gamma - \tilde{\kappa}/2)) + i\delta(k)\Gamma - \tilde{\kappa}/2(\Gamma - \tilde{\kappa}/2) - g^2}{\delta(k)^2 + \tilde{\chi}(\delta(k) + i(\Gamma - \tilde{\kappa}/2)) + i\delta(k)(\kappa + \Gamma) - (\kappa + \tilde{\kappa}/2)(\Gamma - \tilde{\kappa}/2) - g^2}
\end{equation}
and
\begin{equation}
\label{eq:reflection_g_dc}
\tilde{r}(k) = \frac{-i\kappa\delta(k) + \kappa(\Gamma - \tilde{\kappa}/2)}{\delta(k)^2 + \tilde{\chi}(\delta(k) + i(\Gamma - \tilde{\kappa}/2)) + i\delta(k)(\kappa + \Gamma) - (\kappa + \tilde{\kappa}/2)(\Gamma - \tilde{\kappa}/2) - g^2},
\end{equation} 
with an addition of $\tilde{\kappa}/2$ to $\gamma_g/2$ in the exponential decay of the amplitudes. Similarly when the atom is in the state $\ket{f}$ the new coefficients are
\begin{equation}
\label{eq:reflection_f_dc}
\tilde{r}'(k) = \frac{\delta(k) - \tilde{\chi} - i\tilde{\kappa}/2}{\delta(k) - \tilde{\chi} + i(\kappa - \tilde{\kappa}/2)}
\end{equation}
and
\begin{equation}
\label{eq:transmission_f_dc}
\tilde{t}'(k) = -\frac{i\kappa}{\delta(k) - \tilde{\chi} + i(\kappa - \tilde{\kappa}/2)},
\end{equation}
without any change in the exponential decay of the amplitudes.

From these results it is straightforward, following what was done above, to get the new formula for the error which is
\begin{equation}
P_{err} \approx \frac{1}{4}\bigg((\gamma_g+\gamma_f+\tilde{\kappa})n\sigma_T + \frac{\kappa(\Gamma-\tilde{\kappa}/2)}{g^2} + \frac{1}{2\sigma_T^2\kappa^2}\bigg(1 - \frac{\kappa^2}{g^2}\bigg)^2\bigg).
\end{equation}
We notice that $\tilde{\chi}$ has no effect on the error.

\section{First step error minimization}
\label{sec:min}
In this section we will estimate the achievable performance of our scheme for realistic superconducting artificial atoms. Above
we derived the error rate for the case in which the transition between the two lower states is allowed but far from resonance $\Delta \gg g_{gf}$. For a nearly harmonic system we can relate the coupling constants of the two transitions by $g = \sqrt{2}g_{gf}$. Adding a higher order term from the expansion, which will be important below, we arrive at the expression for the error.
\begin{equation}
\label{eq:errror_formula}
P_{err} \approx \frac{1}{4}\bigg({\left(\gamma_g+\gamma_f+\frac{g^2\kappa}{\Delta^2}\right)}n\sigma_T + \frac{\kappa(\Gamma-\tilde{\kappa}/2)}{g^2} + \frac{1}{2\sigma_T^2\kappa^2}\bigg(1 - \frac{\kappa^2}{g^2}\bigg)^2+ \frac{15}{32(g\sigma_T)^{6}}\bigg).
\end{equation}
For simplicity we here only include a simplified expression for the sixth order term valid for $\kappa=g$. The inclusion of this term allows us to get simple expressions for the optimal performance, but for a full agreement with the results of our numerical simulations we need to include a more complicated expression for the sixth order term as well as a fourth order term which is non-vanishing for $\kappa\neq g$.

We shall now optimize the above expression to find the ideal performance for a given atom. A full optimization is beyond the scope of this article and we shall only make a rough analytical optimization. For typical parameters we find that the imperfection arising from the $\Gamma$ term will be less important for us, and we shall therefore ignore it.
First we consider the error from the cavity induced decay rate $g^2\kappa n\sigma_T/\Delta^2$. For typical experimental values this term would be detrimental for the performance of the transistor. We therefore assume that we work with a reduced coupling constant, e.g., by placing the atom near an antinode of the field. A low value of $g$ reduces the error from the cavity induced decay, but this comes at the expense of larger errors from the scattering dynamics. To optimize the performance we therefore assume that $g$ is reduced such that the cavity induced decay rate matches the inherent decay rate in the system. For $\kappa=g$ this gives us
\begin{equation}
g^{opt} = \big(\Delta^2(\gamma_g+\gamma_f)\big)^{1/3}.
\end{equation}

We then consider the optimal pulse width $\sigma_T$. It is evident from the expression above that the condition $g = \kappa$ is particularly favorable for the reduction of the error. This condition ensures that the dispersion relations near resonance are the same regardless of whether the atom is in state $\ket{g}$ or $\ket{f}$. This minimizes the error since it ensures a maximal interference of the outgoing wave packets. For $g = \kappa$ the leading order term in the expansion in the pulse width $\sigma_T$ vanishes. To find the limitation in this situation we therefore need to include more terms in the expansion. For $g = \kappa$ we find that the lowest non-vanishing term is the sixth order term included in Eq \eqref{eq:errror_formula}. In reality one may, however, not be able to fulfill the condition $g = \kappa$ exactly and we therefore parametrize potential deviations from this limit by a parameter $\alpha$ defined by $\kappa= (1+\alpha)g$. 

There are then two different limits for the optimization. The first is when $\alpha$ is sufficiently small that we are limited by the sixth order term. In this case we ignore the error from having a finite $\alpha$. Setting the derivative of the error with respect to $\sigma_T$ equal to zero gives
\begin{equation}
{\sigma_T}_1 = \bigg(\frac{45}{32n\Delta^4(\gamma_g+\gamma_f)^3}\bigg)^{1/7}.
\end{equation}
With this pulse duration we then find an error given by 
\begin{equation}
P_{err}^{opt,1}= \frac{77}{264}(180n^6)^{1/7}\bigg(\frac{\gamma_g+\gamma_f}{\Delta}\bigg)^{4/7}.
\end{equation}
In the opposite case in which $\alpha$ is large enough that we are dominated by the second order term we ignore the sixth order term and find that the optimal pulse duration is given by
\begin{equation}
{\sigma_T}_2 = \bigg(\frac{3\alpha^2}{n(1+\alpha)^2\Delta^{4/3}(\gamma_g+\gamma_f)^{5/3}}\bigg)^{1/3},
\end{equation}
resulting in an error 
\begin{equation}
P_{err}^{opt,2}= \frac{2}{3^{2/3}}(n\alpha)^{2/3}\bigg(\frac{\gamma_g+\gamma_f}{\Delta}\bigg)^{4/9}.
\end{equation}
We can then approximate the true optimum by the larger of the two limits identified here
\begin{equation}
P_{err}^{opt} = \text{Max}[P_{err}^{opt,1},P_{err}^{opt,2}].
\end{equation}

To insert concrete numbers we consider transmons confined in three dimensional cavities which have recently shown amazing
coherence times. For instance Ref. \cite{rigetti} has observed $T_1 = 70\mu$s and $T^*_2 = 92\mu$s for the lower transition. With these numbers we obtain a relaxation rate $\gamma_{1g}/2\pi = (1/T_1)/2\pi = 2.3$ kHz and a coherence decay rate for the lower two levels $(\gamma_{f} + \gamma_{g})/2\pi = (2/T^*_2)/2\pi = 3.5$ kHz.
Approximating the anharmonicity $\Delta$ by the charging energy $E_c$ (cf. Ref. \cite{koch}) gives $\Delta/2\pi = 206$ MHz. With these numbers, allowing for $\alpha\sim 10\%$, and taking $n=8$, we find that the error is slightly dominated by the sixth order term and is given by $P_{err}\approx 0.9\%$ .
For a different experiment \cite{sears} we have $\Delta/2\pi = 340$ MHz, $T_1 = 28 \mu$s and $T^*_2 = 32 \mu$s. For these parameters we find an error probability $P_{err} \approx 1.2\%$ limited again by the sixth order term.

To verify these predictions we have performed a full numerical simulation of the first step, where we truncate the Gaussian at the end of the period. In Fig. \ref{fig:error} we plot the error probability as function of $\alpha$. In the figure we compare the numerical simulation with an analytical error formula where we include terms of fourth and sixth order in $1/\sigma_T$, not reported explicitly above. Here the fourth order term has a component linear in $\alpha$ which displaces the optimal performance slightly away from $\alpha=0$. The values of the parameters correspond to the values for the experiment of Ref. \cite{sears} as stated above. As seen from the figure the full analytical results are in excellent agreement with the numerical simulation, but the approximately optimized function show some minor discrepancies due to the simplicity of the treatment. 

\begin{figure}
\includegraphics[width=10cm]{error}
\caption{Analytical and numerical error as function of $\alpha = (\kappa - g)/g$. The values of the parameters correspond to the values of Ref. \cite{rigetti} and are $\Delta= (2\pi) 206$ MHz, $\gamma_g = (2\pi)3.5$ kHz, $g=g^{opt}=(2\pi) 5.3$ MHz, $\gamma_f=0$, $\gamma_e=2\gamma_g$, and $n=8$. The pulse length is chosen according to the approximate optimization discussed in the text.}
\label{fig:error}
\end{figure}

Alternatively the procedure we discuss here could also be implemented with transmon qubits not being confined in 2D cavities. In this case Ref. \cite{houck} reports $\Delta/2\pi = 455$ MHz, $T_1 = 1.57 \mu$s, and $T^*_2 = 2.94 \mu$s which leads to $P^{err}\approx 3.3\%$

\section{Multiphoton dynamics}

During the second step we assume the signal field to be in a strong coherent state, i.e. with average photon number much greater than one. We can then assume a semi-classical description of the external fields and replace the term in the Hamiltonian \eqref{eq:hamiltonian} describing the interaction between the waveguides modes and the cavity with the term 
\begin{equation}
H^{CW} = i\sqrt{\kappa}\xi(\hat{a}^\dagger - \hat{a}).
\end{equation}
As opposed to seeing this as a semiclassical approximation it can also be seen as an exact transformation valid for incident coherent states, if one uses the transformation discussed below for the analytical derivation of the reflection.
Here $\xi$ is the amplitude of the external incoming field, which we assume to be real and on resonance with the cavity frequency. We account for the decay of the cavity through the two mirrors into the waveguides by introducing two identical Lindblad operators $\hat{c}_{\kappa} = \sqrt{\kappa}\hat{a}$.
The master equation for the system, neglecting for simplicity the transversal decay rates, is then
\begin{equation}
\label{eq:master_equation_multi}
\dot{\rho} = -\frac{i}{\hbar}\big[H, \rho\big] + \sum_{i= \kappa,\gamma_{e1},\gamma_{g1},..}\big[ - \frac{1}{2}\big\{\rho, \hat{c}^\dagger_i \hat{c}_i\big\} + \hat{c}_i\rho\hat{c}^\dagger_i \big],
\end{equation}
where the Hamiltonian $H$ is now only the Hermitian part of the Hamiltonian in \eqref{eq:hamiltonian}, and $\hat{c}_i$ are the Lindblad operators which account for the six decays. 

We are interested in the transmission probability if the atom is initially in state $\ket{g}$. (In the other case the scattering dynamics is equivalent to the first part of the protocol). For simplicity we assume a steady state approximation, which means that we consider a time $T' \gg 1/\kappa$, but assume $T' \ll 1/\gamma_{1g}$ so that we can ignore the decay of state $\ket{g}$ to $\ket{f}$ and set $\gamma_{g1} = 0$. We also ignore all the dephasing operators keeping only the decay from the excited state. With this approximation the Hilbert space reduces to a tensor product of the two-dimensional atomic space spanned by $\ket{g}$ and $\ket{e}$ and the Fock space of the cavity mode excitations. The master equation \eqref{eq:master_equation_multi} is then equivalent to the system of equations
\begin{equation}
\begin{split}
\dot{\rho}_{n g, n g} = +\gamma_{e1} {\rho}_{n e, n e} - 2\kappa n {\rho}_{n g, n g} + 2\kappa(n+1){\rho}_{(n+1) g,(n+1) g} + \\
+ g\sqrt{n} D_{(n-1) e, n g} + \sqrt{\kappa}\xi D_{(n-1) g, n g} - \sqrt{\kappa}\xi D_{(n+1) g, n g} \\
\dot{\rho}_{n e, n e} = -\gamma_{e1} {\rho}_{n e, n e} - 2\kappa n {\rho}_{n g, n g} + 2\kappa(n+1){\rho}_{(n+1) g, (n+1) g} + \\
- g\sqrt{n} D_{(n+1) g, n e} + \sqrt{\kappa}\xi D_{(n-1) e, n e} - \sqrt{\kappa}\xi D_{(n+1) e, n e} 
\end{split}
\end{equation}
where we have defined $D_{a, b} = \rho_{a, b} + \rho_{b, a}$, which have to satisfy 
\begin{equation}
\begin{split}
\dot{D}_{n g, m g} = + \gamma_{e1}{D}_{n e, m e} - \kappa(n+m){D}_{n g, m g} + 2\kappa\sqrt{n+1}\sqrt{m+1}{D}_{(n+1) g, (m+1) g} +\\
+ g(\sqrt{n}{D}_{(n-1) e, m g} + \sqrt{m}{D}_{n g, (m-1) e}) + \sqrt{\kappa}\xi(\sqrt{n}{D}_{(n-1) g, m g} + \sqrt{m}{D}_{n g, (m-1) g} - \\ - \sqrt{n+1}{D}_{(n+1) g, m g} - \sqrt{m+1}{D}_{n g, (m+1) g}), \\
\dot{D}_{n e, m e} = - \gamma_{e1}{D}_{n e, m e} - \kappa(n+m){D}_{n e, m e} + 2\kappa\sqrt{n+1}\sqrt{m+1}{D}_{(n+1) e, (m+1) e} +\\
- g(\sqrt{n+1}{D}_{(n+1) g, m e} + \sqrt{m+1}{D}_{n e, (m+1) g}) + \sqrt{\kappa}\xi(\sqrt{n}{D}_{(n-1) e, m e} + \sqrt{m}{D}_{n e, (m-1) e} - \\ - \sqrt{n+1}{D}_{(n+1) e, m e} - \sqrt{m+1}{D}_{n e, (m+1) e}). \\
\dot{D}_{n e, m g} = - \gamma_{e1}/2{D}_{n e, m e} - \kappa(n+m){D}_{n e, m g} + 2\kappa\sqrt{n+1}\sqrt{m+1}{D}_{(n+1) e, (m+1) g} +\\
- g\sqrt{n+1}{D}_{(n+1) g, m g} + g\sqrt{m+1}{D}_{n e, (m-1) e} + \sqrt{\kappa}\xi(\sqrt{n}{D}_{(n-1) e, m g} + \sqrt{m}{D}_{n e, (m-1) g} - \\ - \sqrt{n+1}{D}_{(n+1) e, m g} - \sqrt{m+1}{D}_{n e, (m+1) g}).
\end{split}
\end{equation}
% I DON'T THINK WE NEED THE NEXT COMMENT SINCE WE CAN JUST SWITCH n and m? and similarly for $D_{n g, m g}$. 
This is an infinite set of equations, but for every given field amplitude $\xi$ we can truncate the Hilbert space at a finite number of excitations $n'$. %The number of states is then $2n'+1$ so that the number of independent elements of the density matrix, which is also the dimension of the linear problem, is $d = (2n'+1)(n'+1)$.
The equations can then be solved by a computer for reasonable values of $n'$. 

The transmitted intensity is given by
\begin{equation}
I_T = \kappa \braket{\hat{a}^\dagger\hat{a}}_{SS} = \kappa \text{Tr}[\hat{a}^\dagger\hat{a}\rho_{SS}] = \kappa \sum_n n\,\rho_{n,n}
\end{equation}
where $\rho_{n, n} = \rho_{n g, n g} + \rho_{n e, n e}$. Similarly the intensity lost because of the decay of the excited state in modes different from the cavity is 
\begin{equation}
I_L = \gamma_e \braket{\sigma_{ee}}_{SS} = \gamma_e \text{Tr}[\sigma_{ee}\rho_{SS}] = \gamma_e \sum_n \rho_{n e,n e}.
\end{equation}
The reflected intensity is given simply by the relation $I_R = I_{IN} - I_T - I_L$.
From these results we produce the curves shown in Fig. \ref{fig:reflection} of the main paper. 

To find an analytical expression for the reflected intensity for strong incident fields it is convenient to change to the Heisenberg picture. The Heisenberg equation of motion for the operator $\hat{a}(t)$ is
\begin{equation}
\label{eq:Heisenberg_equation}
\frac{d}{dt}\hat{a}(t) = -\kappa\hat{a}(t) + g\sigma_{eg}^-(t) + \sqrt{\kappa}\hat{b}_{IN},
\end{equation}
where we have performed the Markov approximation and defined the operator
\begin{equation}
\label{eq:input_field_2}
\hat{b}_{IN}(t) = \frac{1}{\sqrt{2\pi}}\int^{+\infty}_{-\infty} dk\,\hat{b}_0(k)e^{-i(ck-\omega_c)(t-t_0)},
\end{equation}
 in a manner analogous to what was done for the single photon in the Schr\"odinger picture. 
Since the input field is in a strong coherent state we can rewrite Eq. \eqref{eq:Heisenberg_equation} by displacing the operator $\hat{b}_{IN}(t)$ with a constant classical field $\xi$ which we assume to be real: $\hat{b}_{IN}(t)\rightarrow \xi+\hat{b}_{IN}(t)$, where $\hat{b}_{IN}(t)$ is now in the vacuum state \cite{mollow}. With this replacement we then get the Heisenberg equation of motion 
\begin{equation}
\label{eq:Heisenberg_equation_2}
\frac{d}{dt}\hat{a}(t) = -\kappa {\left(\hat{a}(t) - \frac{\xi}{\sqrt{\kappa}}\right)} + g\sigma_{eg}^-(t),
\end{equation}
where we have for simplicity ignored the incoming vacuum operator $\hat{b}_{IN}(t)$ since it will not contribute to the intensity calculated below.

We are interested in describing the strongly driven cavity for which the interaction with the atom is a perturbation to the cavity field. For this purpose it is therefore convenient to look at the operator describing the deviation of the field from the cavity without the atom. This is given by 
\begin{equation}
\hat{a}'(t) = \hat{a}(t) - \frac{\xi}{\sqrt{\kappa}},
\end{equation}
and has the equation of motion 
\begin{equation}
\label{eq:Heisenberg_equation_3}
\frac{d}{dt}\hat{a}'(t) = -\kappa \hat{a}'(t) + g\sigma_{eg}^-(t).
\end{equation}
Since we have transformed away the incident field we are left with a set of equations describing a cavity without any input field. The incident field only appears as a classical driving term on the atom. This can be found by making the replacement $\hat{a}(t) = \hat{a}'(t) + \xi/\sqrt{\kappa}$ in the interaction Hamiltonian, which then results in a driving term 
\begin{equation}
H=\frac{\xi g}{\sqrt{\kappa}} (\sigma_{eg}^-+\sigma_{eg}^+).
\label{eq:rabi}
\end{equation}
Since the cavity has no reflection in the absence of the atom, the reflected intensity is only given by the operator $\hat{a}'(t)$, and can be expressed by
\begin{equation}
I_R = \kappa\braket{\hat{a}'^\dagger\hat{a}'}.
\end{equation}
To find the reflected intensity we formally integrate Eq. \eqref{eq:Heisenberg_equation_3}, to get
\begin{equation}
\label{eq:Heisenberg_equation_4}
\hat{a}'(t) = g\int^{t}_{-\infty}dt' e^{-\kappa(t-t')}\sigma_{eg}^-(t').
\end{equation}
Inserting this and the equivalent for $\hat{a}'^\dagger$ we then get
\begin{equation}
I_R = \kappa g^2 \int^{t}_{-\infty}dt' \int^{t}_{-\infty}dt'' e^{-\kappa(2t-t'-t')}\braket{\sigma_{eg}^+(t'')\sigma_{eg}^-(t')}. 
\end{equation}
We can thus find the reflected intensity from the two point correlation function of the atom. 

We now change the coordinates of the integration by introducing $\tau = t''-t'$ and $T = t'+t''$. Paying attention to the limits of integration we can rewrite the last equation as
\begin{equation}
I_R = \frac{\kappa g^2 }{2}\int^{+\infty}_{-\infty}d\tau \int^{2t-|\tau|}_{-\infty}dT e^{-\kappa(2t-T)}\braket{\sigma_{eg}^+(\tau)\sigma_{eg}^-(0)}, 
\end{equation}
where we have used the fact that in steady state $\braket{\sigma_{eg}^+(t'')\sigma_{eg}^-(t')}$ depends only on the time difference $t''-t'$. Performing the integral over $T$ we get
\begin{equation}
\label{eq:intensity_reflected}
I_R = \frac{g^2}{2} \int^{+\infty}_{-\infty}d\tau e^{-\kappa|\tau|}\braket{\sigma_{eg}^+(\tau)\sigma_{eg}^-(0)}.
\end{equation}
Note that this is an exact result for the steady state reflection, valid as long as the incident field is in a coherent state. To evaluate $\braket{\sigma_{eg}^+(\tau)\sigma_{eg}^-(0)}$ we now restrict ourselves to the limit of strong driving. In steady state and for strong driving there is an equal probability that the atom is in the states $\ket{g}$ and $\ket{e}$ as described by the density matrix
\begin{equation}
\hat \rho_{SS}=\frac{1}{2}
{\left(\ket{g}\bra{g}+\ket{e}\bra{e}\right)}.
\end{equation}
Hence from the application of $\sigma_{eg}^-(0)$ to $\hat \rho_{SS}$ we immediately get a factor $1/2$ since the $\ket{g}\bra{g}$ term vanishes, and the expression is reduced to $\braket{\sigma_{eg}^+(\tau)\sigma_{eg}^-(0)}={\rm Tr}(\sigma_{eg}^+(\tau)\ \ket{g}\bra{e}(0))/2$. This result immediately shows that the steady state reflection is limited from above by $I_R\leq g^2/2\kappa$. To get more than an upper limit we need to consider the  time integral in Eq. \eqref{eq:intensity_reflected}. For strong driving this integral has support over a long time scale compared to the fast oscillations induced by the driving \eqref{eq:rabi}. The rapid driving will essentially decouple the atom from the cavity and for weak decoherence compared to the integration period $\sim1/\kappa$, i.e., $\gamma_e\ll\kappa$,  we can replace $\sigma_{eg}^+(\tau)$ by the time evolution obtained from Eq. \eqref{eq:rabi}. Doing this and averaging over time we get a second factor of 1/2 and we have $\braket{\sigma_{eg}^+(\tau)\sigma_{eg}^-(0)} = 1/4$ after the time average. Inserting this in Eq. \eqref{eq:intensity_reflected} and doing the last integral, we get $I_R = g^2/4\kappa$. For $g\lesssim \kappa$ this asymptotic form requires that the Rabi oscillations are much faster than the cavity decay $\xi g/\sqrt{\kappa}\gg\kappa$ (this result can be understood as the sideband of the Mollow triplet moving outside the cavity bandwidth). This condition translates into $I_{IN}\gg\kappa^3/g^2$, which translates into a very slow convergence for $I_R$ for $\kappa\gg g$ in Fig. \ref{fig:reflection} of the main paper where the $I_{IN}$-axis is rescaled by $g^2/\kappa$. We thus find $I_R$ to be in the range  $I_R = g^2/4\kappa$ to $I_R = g^2/2\kappa$, but take the lower value to be conservative.

Combining this reflected intensity with the fact that we are limited by the decay of the lower transition we find that the gain is given by $G\sim g^2/\kappa(\gamma_{1g}+\tilde\kappa$), when we include the cavity induced decay of the lower transition. To evaluate the possible gain we set $\kappa=g$ and use the optimal value of $g$ derived in Sec. \ref{sec:min}. We then find that the possible gains are $G= 783$, $570$, and $116$ for the parameters of Refs. \cite{rigetti}, \cite{sears}, and \cite{houck} respectively.

% \bibliography{transistor}